\pdfoutput=1
\documentclass[double]{raa_twocolumn}

\usepackage{graphicx,times}
\usepackage{natbib}
\usepackage{amssymb,amsmath}
\bibpunct{(}{)}{;}{a}{}{,}
\usepackage{multirow}

\usepackage{CJKutf8}

\usepackage[pagebackref=true]{hyperref}

\usepackage{tikz}
\usetikzlibrary{positioning,matrix,shapes.arrows}
\usetikzlibrary{arrows,shapes.geometric}
\tikzstyle{startstop} = [rectangle, rounded corners, minimum width=3cm, minimum height=1cm,text centered, draw=black, fill=red!30]
\tikzstyle{io} = [trapezium, trapezium left angle=75, trapezium right angle=105, minimum width=2cm, minimum height=1cm, text centered, draw=black, fill=blue!30]
\tikzstyle{process} = [rectangle, minimum width=3cm, minimum height=1cm, text centered, draw=black, fill=orange!30]
\tikzstyle{decision} = [diamond, minimum width=3cm, minimum height=1cm, text centered, draw=black, fill=green!30]
\tikzstyle{arrow} = [thick,->,>=stealth]

\begin{document}

   \title{Multi-Layer Perceptron for Predicting Galaxy Parameters (MLP-GaP): stellar masses and star formation rates}

 \volnopage{ {\bf 20XX} Vol.\ {\bf X} No. {\bf XX}, 000--000}
   \setcounter{page}{1}

   \author{Xiaotong Guo \begin{CJK*}{UTF8}{gkai}(郭晓通)\end{CJK*}\inst{1}, Guanwen Fang \inst{1}, Haicheng Feng \inst{2}, Rui Zhang \inst{1}
   }

   \institute{ Institute of Astronomy and Astrophysics, Anqing Normal University, Anqing, Anhui 246133, China;{\it guoxiaotong@aqnu.edu.cn} \thanks{guoxiaotong@aqnu.edu.cn}
        \and
             Yunnan Observatories, Chinese Academy of Sciences, Kunming 650011, China\\
\vs \no
   {\small Received 2024 October 1; accepted 2024 October 30}
}

\abstract{The large-scale imaging survey will produce massive photometric data in multi-bands for billions of galaxies. Defining strategies to quickly and efficiently extract useful physical information from this data is mandatory. Among the stellar population parameters for galaxies, their stellar masses and star formation rates (SFRs) are the most fundamental. 
We develop a novel tool, \textit{Multi-Layer Perceptron for Predicting Galaxy Parameters} (MLP-GaP), that uses a machine-learning (ML) algorithm to accurately and efficiently derive the stellar masses and SFRs from multi-band catalogs. 
We first adopt a mock dataset generated by the \textit{Code Investigating GALaxy Emission} (CIGALE) for training and testing datasets. 
Subsequently, we used a multi-layer perceptron model to build MLP-GaP and effectively trained it with the training dataset.
The results of the test performed on the mock dataset show that MLP-GaP can accurately predict the reference values. 
Besides MLP-GaP has a significantly faster processing speed than CIGALE. To demonstrate the science-readiness of the MLP-GaP, we also apply it to a real data sample and compare the stellar masses and SFRs with CIGALE. Overall, the predicted values of MLP-GaP show a very good consistency with the estimated values derived from SED fitting.
Therefore, the capability of MLP-GaP to rapidly and accurately predict stellar masses and SFRs makes it particularly well-suited for analyzing huge amounts of galaxies in the era of large sky surveys.
}

   \authorrunning{X. Guo et al. }            
   \titlerunning{MLP-GaP: stellar masses and SFRs}  
   \maketitle

%
\section{Introduction}
Galaxies are fundamental building blocks of the cosmic large-scale structures and have played a crucial role in the evolution of baryons in the universe's history. We are entering the season of the Stage-IV all-sky surveys, which will observe billions of galaxies. Large sky surveys have become a fundamental tool for cosmology and galaxy formation studies over the past few decades. In particular, Stage-III surveys have provided multi-band data for tens to hundreds of millions of galaxies and other celestial objects, such as the Kilo-Degree Survey \citep[KiDS,][]{2024A&A...686A.170W}, Hyper Suprime-Cam \citep[HSC,][]{2018PASJ...70S...4A}, and the Dark Energy Survey \citep[DES,][]{2021ApJS..255...20A}.
Over the past decade, these large-scale surveys have definitely pushed our understanding of the dark matter distribution in the universe via weak lensing \citep[e.g.,][]{2011arXiv1110.3193L, 2017MNRAS.465.1454H, 2018PhRvD..98d3526A, 2021A&A...646A.129J, 2021A&A...646A.140H, 2023JCAP...09..004M}. 
However, they have also provided useful data to significantly grow our understanding of the formation and evolution of galaxies \citep[e.g.,][]{2018MNRAS.480.1057R,2018ApJ...857..104G,2018PASJ...70S..37G,2021ApJ...923...37A,2023SCPMA..6629513X}. 
In this context, the ability to collect accurate stellar population properties of all examined galaxies, despite being crucial \citep[e.g.,][]{2019A&A...632A..34W}, has been a bottleneck in the science outcomes \citep[e.g.,][]{2021A&A...653A..82B}.

In the next decade, Stage-IV surveys will observe billions of galaxies, providing data that is both more in-depth and of higher quality and covering the wavelengths from ultraviolet (UV) to near-infrared (NIR). 
Observational programs for the Stage-IV surveys include Euclid \citep{2011arXiv1110.3193L}, \textit{Vera Rubin} Legacy Survey in Space and Time \citep[VR/LSST,][]{2019ApJ...873..111I}, and the China Space Station Telescope \citep[CSST, e.g.,][]{2011SSPMA..41.1441Z, 2018RPPh...81f6901Z, 2019ApJ...883..203G}.
The data for Stage-IV surveys will offer unprecedented insights into cosmology and galaxy evolution, including crucial revelations on dark matter, dark energy, and the formation and evolution of galaxies.

Galaxies are complex systems with numerous physical parameters, such as stellar mass (M$_\star$), size, morphology, star formation rate (SFR), age, metallicity and chemical composition. Accurately measuring these parameters is essential for understanding the formation and evolution of galaxies and their role in shaping the structure and evolution of the universe. 
However, obtaining unbiased stellar population parameters for galaxies remains a significant challenge due to the well-known age/metallicity degeneracies. Currently, there is no consensus on how to mitigate these degeneracies, and various attempts have been made to account for different phases of stellar evolution \citep[e.g.,][]{2005MNRAS.362..799M, 2016MNRAS.463.3409V} to solve this problem. On the other hand, combining different codes and stellar libraries, along with stellar population priors and star formation histories, can help reduce the impact of systematic errors \cite[e.g.,][]{2023SCPMA..6629513X}. This approach, though, is time consuming and new strategies are needed to reduce the computational times for testing the largest variety of stellar population models and priors, especially over large datasets. 

Among the different stellar population parameters for galaxies, M$_\star$ and SFR stand out as the most crucial ones. In particular, they have a rather tight relation, the so-called {\it main sequence} of galaxies \citep[e.g.][]{2004MNRAS.351.1151B,2007ApJ...660L..43N,2007A&A...468...33E,2007ApJ...670..156D,2015A&A...575A..74S}, that is a fundamental diagnostic for galaxy formation theories\citep[e.g.,][]{2015MNRAS.450.4486F,2019MNRAS.485.4817D,2023MNRAS.519.1526P}.
Nevertheless, estimating the M$_\star$ and SFR is also a complex task \citep{2012ARA&A..50..531K} as they are directly or indirectly related to the observations of stars. 
M$_\star$ represents the total mass of stars within a galaxy. Low-mass, non-ionizing old stars are the most abundant within a galaxy and contribute significantly to its optical luminosities. Consequently, M$_\star$ is closely associated with its optical luminosity.
Moreover,  the estimation of M$_\star$ also depends on the stellar population model \citep[e.g.,][]{2003MNRAS.344.1000B,2005MNRAS.362..799M} and the Initial Mass Function \citep[IMF; e.g.,][]{1955ApJ...121..161S,2003PASP..115..763C}, and the form of the star formation history (SFH), for which there is no consensus on the impact of systematics \citep[e.g.,][]{10.1093/mnras/stt1280}. 
The SFR represents the rate at which new stars are being formed within a galaxy. It is closely tied to the presence of young stars and the ionized gas that envelops massive stars. The emission from young and massive O/B-type stars is predominantly in the UV band. As a result, UV emission can serve as a valuable indicator of the SFR \citep[e.g.,][]{2011ApJ...741..124H,2012ARA&A..50..531K}. Additionally, emission lines (such as H$\alpha$) from ionized gas can be observed in the optical and NIR bands, further tracing the SFR \cite[e.g.,][]{2002A&A...383..801B,2007ApJS..173..256T}. 
Dust also plays an important role, as this is heavily produced around new stars. This dust absorbs approximately half of the starlight and re-emits it in the far-IR, meaning that far-IR luminosities can also trace the SFR \citep[e.g.,][]{1998ApJ...508..123F,2020ARA&A..58..529S}.

Traditionally, the quantities M$_\star$ and SFR have been estimated primarily through techniques like optical spectroscopy, which involves fitting theoretical models to observed data. For instance, the Sloan Digital Sky Survey (SDSS) MPA--JHU Data Release 8 (DR8) catalog provides stellar masses and SFRs for 1 843 200 galaxies \citep{2003MNRAS.341...33K,2004MNRAS.351.1151B}. However, many surveys lack spectroscopic observations and only provide photometry data, such as KiDS \citep[e.g.,][]{2024A&A...686A.170W}. Despite this, it is still feasible to derive the M$_\star$ and SFR by using their spectral energy distribution (SED) constructed from multi-band photometric data.  
For example, \cite{2019RAA....19...39G} used SED fitting to obtain the SFRs and stellar masses of 145 635 galaxies in the Hawaii-Hubble Deep Field-North.
When we look towards the future with the expected release of multi-band photometric data for billions of galaxies, as there is no way to solve the degeneracies among all the stellar population parameters, the only approach we have is to derive M$_\star$ and SFR using different set-up \citep[e.g.,][]{2023SCPMA..6629513X}, make this dramatically time-consuming.
Given the immense volume of data that will be available, developing a highly efficient method is crucial for extracting meaningful information from these datasets. 

In recent years, the rapid development of machine-learning (ML) algorithms has brought revolutionary changes to various fields, including astronomy. ML has become an integral part of astronomical research, being widely used for a variety of classifications, including astronomical object categorization \citep[e.g.,][]{2024MNRAS.527.4677Z}, galaxy morphology classification \citep[e.g.,][]{2023ApJS..268...34D,2023AJ....165...35F,2024ApJS..272...42S}, and much more. Moreover, ML is also used to predict various parameters and properties of astronomical objects. 
\cite{2022A&A...666A..85L} has developed an innovative ML tool (called GaZNet) capable of predicting galaxy redshifts by integrating both image data and multi-band photometric information.  \cite{2019MNRAS.484.4683W} have trained a deep residual convolutional neural network to predict the gas-phase metallicity of galaxies using three-band \textit{gri} images from the SDSS. \cite{2019A&A...622A.137B} have used the Random Forest to estimate the stellar masses and SFRs of galaxies at redshifts in the range 0.01 < z < 0.3.  When compared to traditional methods, ML not only offers enhanced efficiency but also delivers accuracy. Therefore, to efficiently and accurately derive stellar population parameters for billions of galaxies, we plan to develop an ML algorithm called \textit{Multi-Layer Perceptron for Predicting Galaxy Parameters} (MLP-GaP),  which first is used to estimate the stellar masses and SFRs for galaxies with redshift $z < 3$ in this work.



The structure of this work is as follows.
Section~\ref{sec:data} presents the data used in this work and the process of generating a mock catalog.
Section~\ref{sec:model} describes the ML model used to build the MLP-GaP and outlines its training process.
In Section~\ref{sec:evaluation}, we comprehensively evaluate the performance of the MLP-GaP in the testing dataset, offering a comparative analysis between MLP-GaP and traditional SED fitting techniques.
In Section~\ref{sec:discussion}, we discuss the performance of the MLP-GaP on observational data of actual galaxies and provide its possible future application scenarios, as well as subsequent improvement methods for the algorithm.
Finally, a brief summary is presented in Section~\ref{sec:summary}.



\section{Data}\label{sec:data}
In this work, we want to develop an ML algorithm that uses multi-band photometric data (i.e. aperture fluxes or magnitudes) to predict the stellar population parameters of a given galaxy dataset, similar to the standard SED fitting techniques \citep[e.g.,][]{2019A&A...622A.103B,2006A&A...457..841I}. 
To train and test such a ``supervised'' algorithm (see Section~\ref{sec:model}), it is essential to dispose of a galaxy catalog with ``ground truth'' stellar population parameters. Unfortunately, unlike other galaxy parameters that are straightforward to measure, the stellar population parameters of galaxies can only be indirectly derived from measured parameters. stellar population parameters of galaxies can not obtain ``ground truth'' values. 
However, the derivation of these parameters relies on various assumptions, models, and inherent uncertainties, which means that the parameters obtained from different techniques are not uniform \citep[see][]{2023SCPMA..6629513X} and do not represent their ``ground truth'' values.

In absence of real dataset with ``ground truth'' values, a viable strategy is to produce a realistic mock dataset, which include synthetic photometric data of the realistic galaxies and their known stellar population parameters.
To construct such a catalog of galaxies or dataset, we use the \textit{Code Investigating GALaxy Emission} \citep[CIGALE, V2022.1, ][]{2019A&A...622A.103B} to generate a catalog of mock galaxies with known stellar population parameters (see Section~\ref{sec:mock_gal}). CIGALE is an open-source \textit{Python} code designed to analyze the SEDs of galaxies across a wide range of wavelengths, from X-ray to radio. CIGALE models galaxy SEDs by employing composite stellar populations from simple stellar populations combined with highly flexible star formation histories (SFHs), and this approach can provide the flux densities for various bands, M$_\star$, SFR, attenuation, dust luminosity, and many other physical quantities. In particular, M$_\star$ and SFR which can be considered as reference values for developing MLP-GaP. 
For this work, we adopt a delayed SFH model ($SFR \propto t/\tau^2 \times \exp{-t/\tau}$), which is currently popular and aligns well with observational data.
To derive the stellar population spectrum, we use the stellar population synthesis model from \citet[referred to as BC03]{2003MNRAS.344.1000B}, assuming that the IMF adopts \cite{2003PASP..115..763C}. In addition, we also adopt attenuation law \citep{2000ApJ...533..682C}, and dust emission \citep{2014ApJ...784...83D}.
As mentioned, we also use actual observational data as the basis for generating the catalog to ensure that the redshift and luminosity distributions of the mock galaxy sample are well-aligned with observations and to mimic the observation uncertainties on the photometry (i.e., data noise). 

\subsection{Observation data}\label{sec:obs_data}
\begin{figure}
	\centering
	\includegraphics[width=\linewidth]{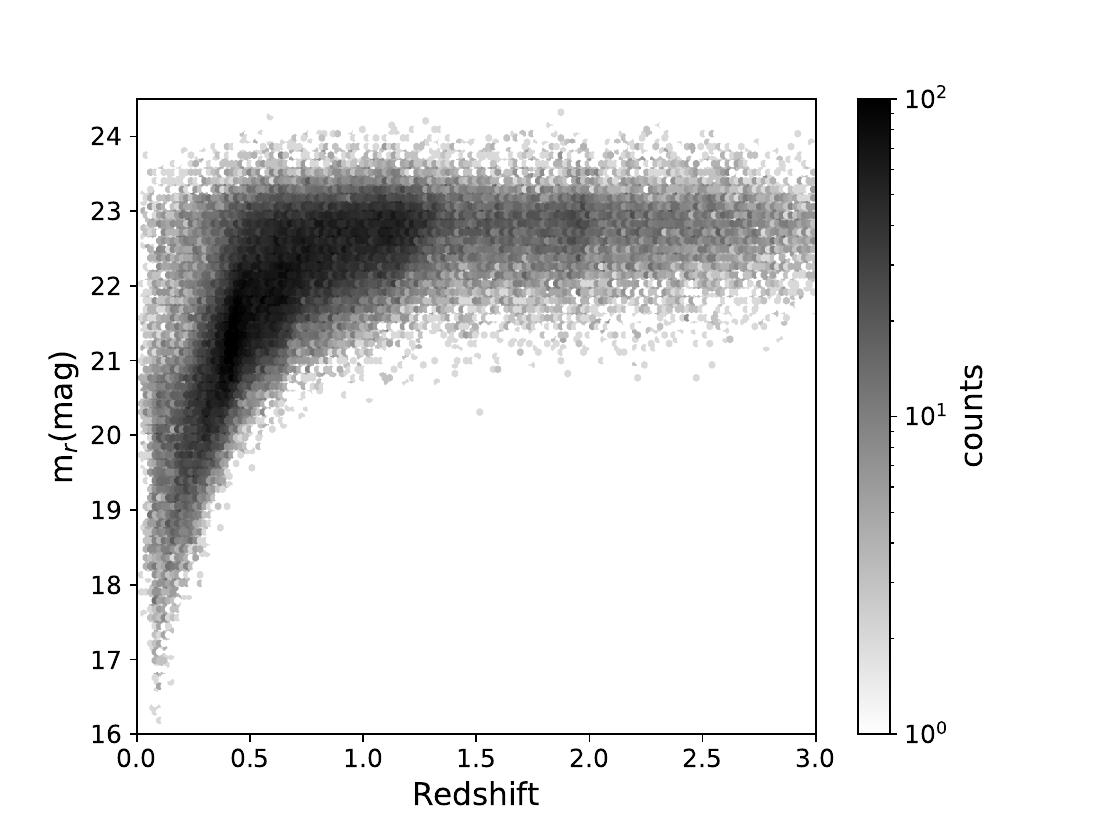}
	\caption{The distribution of r-band magnitude with redshift in real observation data.}
	\label{fig:z-r-mag}
\end{figure}

KiDS provides a unique dataset, with 9-band photometry including 4 optical ($ugri$) and 5 NIR bands (ZYHJK$_s$), to study stellar populations of galaxies among Stage-III surveys, down to a limiting magnitude of $r\sim 24$. MLP-GaP is suitable for application in the KiDS dataset. Therefore, we develop MLP-GaP based on KiDS data.
We firstly have collected redshifts and 9-band magnitudes for 120,000 random galaxies, providing a solid foundation for generating a mock sample that closely aligns with actual observational data. We only used redshifts and r-band magnitudes from actual observation data to generate the mock sample. 
The redshifts used are ``morphoto-z'' obtained by GaZNet \citep{2022A&A...666A..85L}, ranging from 0 to 3. The ``morphoto-z'' is derived by combining imaging and multi-band photometric data. It offers superior accuracy, precision, and fewer outliers than traditional photometric redshifts.
Fig.~\ref{fig:z-r-mag} shows the distribution of redshifts and r-band magnitudes for 120,000 galaxies.
 
\subsection{The catalog for mock galaxies}
\label{sec:mock_gal}
To construct a catalog for mock galaxies, we need to simulate the generation of intrinsic parameters (e.g., age, extinction coefficient) and the observation information (photometry of each band). 
The intrinsic parameters are input parameters for SED fitting, and they are generated randomly within a predefined range (see Table~\ref{tab:parameter}) to emulate the natural variation in galactic properties. 
Among observation information, redshifts and $r$-band photometric data are from actual observation data. 
The aperture photometry from the other bands, their uncertainties, as well as corresponding stellar masses and SFRs are all generated through CIGALE.

\begin{table*}[h]
	\centering
	\caption{The range of SED parameters.\label{tab:parameter}}
	\begin{tabular}{c|c|c|c}
		\hline\hline
		Parameter & Unit & Range & Interval \\
		\hline
		tau & Myr & 250 --- 8000 & 50\\
		age & Myr & 250 --- Cosmology age at the redshift & 50\\
		metallicity & &0.0001, 0.0004, 0.004, 0.008, 0.02, 0.05 & \\
		$A_V$ & mag & 0 --- 2 & 0.01\\
		alpha &   & 0.0625 --- 4 & 0.0625\\
		\hline
	\end{tabular}
\end{table*}

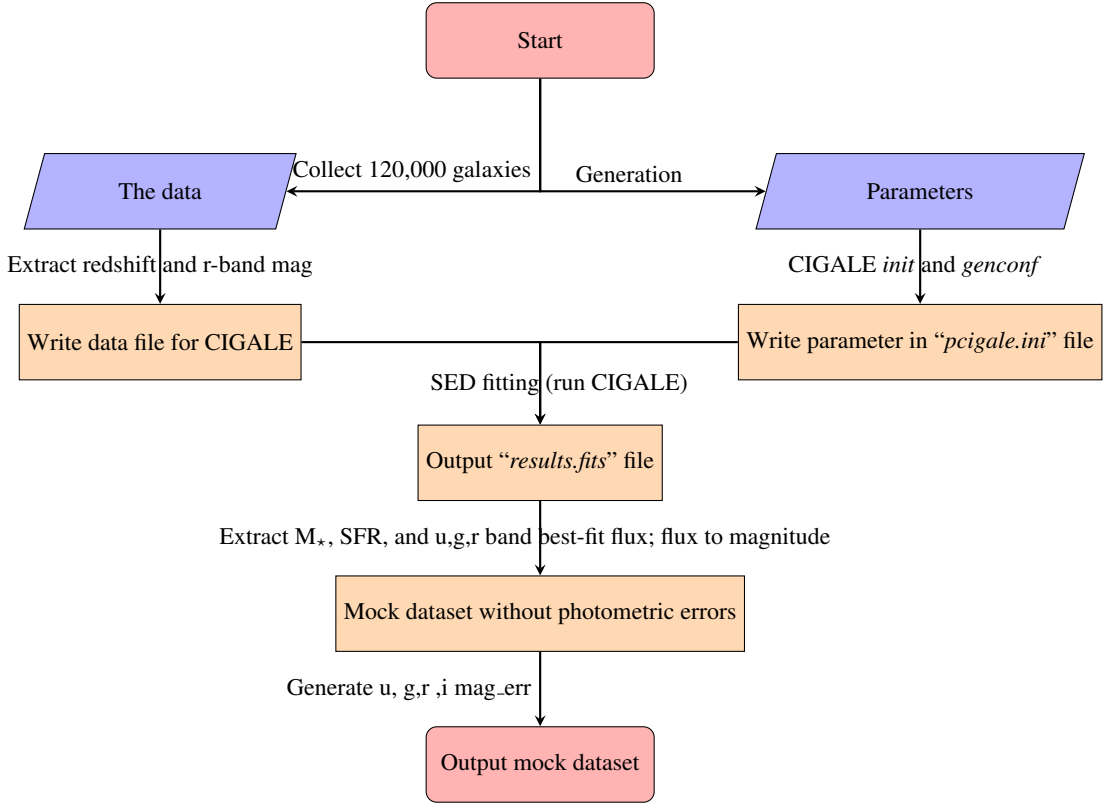
\begin{figure*}
	\centering
	\begin{tikzpicture}[node distance=2cm]
		\node (start) [startstop] {Start};
		\node (in1) [io, below of=start, xshift=-5cm,yshift=0cm] {The data};
		\node (in2) [io, below of=start, xshift=5cm, yshift=0cm] { Parameters};   
		\node (pro1) [process, below of=in1, xshift=0cm, yshift=0cm] {Write data file for CIGALE};
		\node (pro2) [process, below of=in2, xshift=0cm, yshift=0cm] {Write parameter in \textquotedblleft \textit{pcigale.ini}\textquotedblright \ file };
		\node (pro3) [process, below of=start, xshift=0cm, yshift=-3.6cm] {Output \textquotedblleft \textit{results.fits}\textquotedblright \ file};
		\node (pro4) [process, below of=pro3, xshift=0cm, yshift=0cm] {Mock dataset without photometric errors};
		\node (stop) [startstop, below of=pro4, xshift=0cm, yshift=0cm] {Output mock dataset};
		
		\draw [arrow](start) |-node[anchor= south east ] {Collect 120,000 galaxies} (in1);
		\draw [arrow](start) |-node[anchor= south west , xshift=0.35cm] {Generation} (in2);
		\draw [arrow](in1) --node[anchor= center ] {Extract redshift and r-band mag} (pro1);
		\draw [arrow](in2) --node[anchor= center, xshift=-0.07cm ] {CIGALE \textit{init} and \textit{genconf}} (pro2);
		\draw [arrow](pro1) -| node[anchor= north ,xshift=0.25cm,yshift=-0.3cm] {SED fitting (run CIGALE)} (pro3);
		\draw [arrow](pro2) -| (pro3);
		\draw [arrow](pro3) --node[anchor= center ,xshift=-0.2cm] {Extract M$_\star$, SFR, and u,g,r band best-fit flux; flux to magnitude} (pro4);
		\draw [arrow](pro4) -- node[anchor=east] {Generate u, g,r ,i mag\_err} (stop);
	\end{tikzpicture}
	\caption{The flowchar for generating mock dataset for 120,000 galaxies.}
	\label{fig:flowchar}
\end{figure*}

First, we create a data file for each galaxy, which includes observed photometric data and redshift. Among them, the redshift and flux density of the $r$-band are based on actual observation data from 120,000 galaxies. Assuming that the error of the $r$-band magnitude is 0.1 mag, it can convert that into the error of the flux density of the $r$-band. Using placeholder values, such as ``-9999'', represent data from other missing bands. 
Then, initialize CIGALE in the directory where data files are located and generate configuration files.
Next, we modify and update the configuration files. In this process, we write the file names, the specific modules used, and the corresponding parameters (i.e., tau, age, metallicity) into the configuration file. 
The modules and parameters for SED fitting are summarized in Table~\ref{tab:module}. 
Final, running CIGALE, to generate the photometric data for missing bands in the input data file, along with calculating the stellar masses and SFRs.
To have realistic uncertainties on the photometry of the mock galaxies derived as above, we start from the errors reported in the catalog and perform a random sampling within a certain range. Suppose we need to generate a realistic error for the u-band photometry of a mock galaxy with a magnitude of 19.3. We first identify galaxies in the actual galaxy sample with u-band magnitudes close to this value, specifically within a $\pm0.01$ mag range (i.e., from 19.29 to 19.31). If the magnitudes of all galaxies reported in the catalog does not fall within a specified range, we will expand our search to a broader range until we find galaxies that meet the criteria. Then, the error is randomly selected from among these errors. Fig.~\ref{fig:flowchar} presents a flowchart that comprehensively outlines the process for generating the mock catalog for galaxies.

\begin{table*}[h]
	\centering
	\caption{The module assumptions for mock dataset.\label{tab:module}}
	\begin{tabular}{c|c|c}
		\hline\hline
		Module & Parameters & Value \\
		\hline
		\multirow{5}{*}{SFH(delayed)} & Tau\_main(Myr) & tau \\
		& Age(Myr) & age \\
		& f\_burst & 0 \\
		& Tau\_burst(Myr) & 50 \\
		& burst\_Age(Myr) & 100 \\
		\hline
		\multirow{2}{*}{BC03}&IMF&1(Chabrier)\\
		&Metallicity& metallicity\\
		\hline
		&&\\
		dustatt\_modified\_starburst & E\_BV\_lines(mag) & $\frac{A_V}{3.1*0.44}$ \\
		&&\\
		\hline
		dale2014& Alpha& alpha\\
		\hline
	\end{tabular}
	\\
	Notes: tau, age, metallicity, $A_V$, alpha are the parameters generated by the simulation of each galaxy, and their ranges are in Table~\ref{tab:parameter}.
\end{table*}

\subsection{Partition of Datasets}
The mock dataset of 120,000 galaxies, as generated in the previous section, consists of redshifts, 9-band magnitudes, their associated errors, their stellar masses, and SFRs. Within this dataset, the redshifts, 9-band magnitudes, and their associated errors are utilized as the input parameters, or ``features'', for our ML algorithm. Furthermore, the corresponding stellar masses and SFRs are the desired outcomes, or ``targets'', that our ML algorithm aims to predict.

This dataset is used to train and test the MLP-GaP. 
To proceed with the training, validation and testing of the MLP-GaP, it is important to partition the dataset appropriately. 
Hence, we split the dataset of 120,000 mock galaxies into three separate samples:
\begin{itemize}
	\item [$\bullet$]  \textbf{Training dataset}: Consisting of 90,000 galaxies, this dataset is used to train the MLP-GaP.
	\item [$\bullet$]  \textbf{Validation dataset}: Comprising 10,000 galaxies, this dataset is utilized to tune hyperparameters and prevent overfitting during the training process.
	\item [$\bullet$]  \textbf{Testing dataset}: With 20,000 galaxies, this dataset serves as unseen data to assess the MLP-GaP's performance and conduct error statistical analysis.
\end{itemize}
This partitioning strategy comprehensively evaluates the performance of the MLP-GaP while maintaining a balance between training, validation, and testing datasets.


\section{MLP-GaP Architecture and Training} \label{sec:model}
Due to our goal of constructing a mapping between photometric data of galaxies and their parameters, we use supervised ML algorithms. 
Given our algorithm is expected to predict continuous values, which aligns with a regression problem, using multi-layer perceptrons (MLPs) is the best choice. MLPs are a class of feedforward artificial neural networks characterized by their fully connected architecture and the use of nonlinear activation functions. MLPs include at least three layers: an input layer, one or more hidden layers, and an output layer.

\subsection{Architecture}

To accurately predict both M$_\star$ and SFR, MLP-GaP is built using an MLP model with 10 layers. Its architecture can be described as follows: 
\begin{itemize}
	\item [$\bullet$] \textit{Input Layer}: 19 nodes;
	\item [$\bullet$]  \textit{Hidden Layers}: 512, 512, 512, 512, 256, 256, 128, 64, 32 nodes with Rectified Linear Unit (ReLU) activation function;
	\item [$\bullet$]  \textit{Output Layer}: 2 nodes for predicting M$_\star$ and SFR.
\end{itemize}
To enhance our model's performance and mitigate the risk of overfitting, we incorporate both training and validation datasets within our training domain.
Our model uses Huber loss \citep{Friedman99+huberloss} function, which provides a balanced approach for evaluating the performance of a regression model \citep[e.g.,][]{2022ApJ...929..152L}.
The Huber loss is defined as
\begin{equation}
	L_\delta(a) = \left\{
	\begin{aligned}
		&\frac{1}{2} a^2,& |a|\leq\delta \\
		&\delta(|a|-\frac{1}{2}\delta), & \text{otherwise}.\\
	\end{aligned}
	\right .
\end{equation}
Where $a=y_{\text{ture}}-y_{\text{pred}}$, $y_{\text{ture}}$ is the reference values for the simulations, $y_{\text{pred}}$ is the predicted values by the MLP model. $\delta$ is a threshold parameter that can be pre-setted and fixed to 0.001 in this work. Compared to traditional loss functions, the Huber loss function can provide better robustness, effectively mitigate the impact of outliers, maintain sensitivity when errors are small, and exhibit linear characteristics when errors are large, making the optimization process more stable and rapid.
Additionally, we use the Adam optimizer \citep{2014arXiv1412.6980K} to facilitate the optimization process, ensuring efficient and effective model training. 
\subsection{Training}
To enhance the training process and ensure that the model can converge effectively to the global optimum while avoiding entrapment in local optima, we have adopted a segmented training method. This method is more flexible than the decay rate strategy, permitting us to make necessary adjustments to the learning rate at various training stages based on the model's performance. Furthermore, considering that adjustments may need to be made to the model during the training process, the segmented training method provides us with increased control, thereby significantly improving the efficiency and effectiveness of our model training. Next, we will provide a detailed exposition of our training process.

Initially, the model was trained for 20 epochs with a learning rate of $10^{-3}$. Starting with a high learning rate can lead to a faster reduction of the loss and accelerate the convergence of the model by making larger updates to the weights. Then, the learning rate is set to $10^{-4}$, the pre-trained model is re-loaded, and the model is trained for 50 epochs. Next, the learning rate is reduced again to $10^{-5}$, and the model is trained for another 50 epochs. If the model has not fully converged, repeat the previous training until the model is well-trained. Ultimately, our model converged with a loss function value of $7.63\times 10^{-6}$.


\section{Evaluation of MLP-GaP on testing dataset} \label{sec:evaluation}
After building and training the MLP-GaP, to further assess its performance, it will be applied to estimate the stellar masses and SFRs of the galaxies in the testing dataset. Firstly, the predictive values of the MLP-GaP are compared with their reference values. Subsequently, the predictive values of the MLP-GaP are compared with the estimated values of CIGALE.

\subsection{The evaluator metrics}
To assess the performance of the MLP-GaP in terms of accuracy and precision, we use three different statistical estimators:
\begin{enumerate}
	\item The coefficient of determination $R^2$:
	\begin{equation}
		R^2 = 1- \frac{\sum_{i=1}^{N}(y^{i}_{\mathrm{pred}}- y^{i}_{\mathrm{true}})^2}{\sum_{i=1}^{N}(y^{i}_{\mathrm{pred}}- \bar{y}_{\mathrm{true}})^2}.
	\end{equation}
	\item The Mean Absolute Error (MAE):
	\begin{equation}
		\mathrm{MAE} = \frac{1}{N} \sum\limits_{i=1}^{N} \left| y^{i}_{\mathrm{pred}}- y^{i}_{\mathrm{true}} \right|.
	\end{equation}
	\item The Mean Squared Error (MSE):
	\begin{equation}
		\mathrm{MSE} = \frac{1}{N} \sum\limits_{i=1}^{N} \left( y^{i}_{\mathrm{pred}}- y^{i}_{\mathrm{true}} \right)^2.
	\end{equation}
\end{enumerate}
Where $N$ is total number of the data points, $y^{i}_{\mathrm{pred}}$ is the predictive or estimated values, $y^{i}_{\mathrm{true}}$ is the reference values, and
$\bar{y}_{\mathrm{true}}$ is the mean value of the reference values.

$R^2$ is used to measure the goodness of fit of a regression model. 
It ranges from 0 to 1 and is a proportion of the variance in the dependent variable that is predictable from the independent variables. 
In practical applications, $R^2$ values close to 1 are desirable, indicating that the model fits the data well. 
However, a high $R^2$ value does not necessarily mean that the model is accurate or precise, as it only measures the fitting goodness of the model and not the accuracy of its predictions. MAE and MSE measure the average absolute and squared differences between the predicted and actual values. Both MAE and MSE are non-negative values, and lower values for both metrics indicate better performance of the ML model. 
Therefore, the comprehensive evaluation using $R^2$, MAE, and MSE can serve as a better measure of the MLP-GaP's performance. In our analysis, we assess the performance of the MLP-GaP by comparing its predictions to reference values. The closer the $R^2$ value is to 1, and the lower the MAE and MSE values are, the better the performance of the MLP-GaP is considered to be.

\subsection{Comparing the MLP-GaP predictions with reference values}
\begin{figure*}
	\centering
	\includegraphics[width=\linewidth]{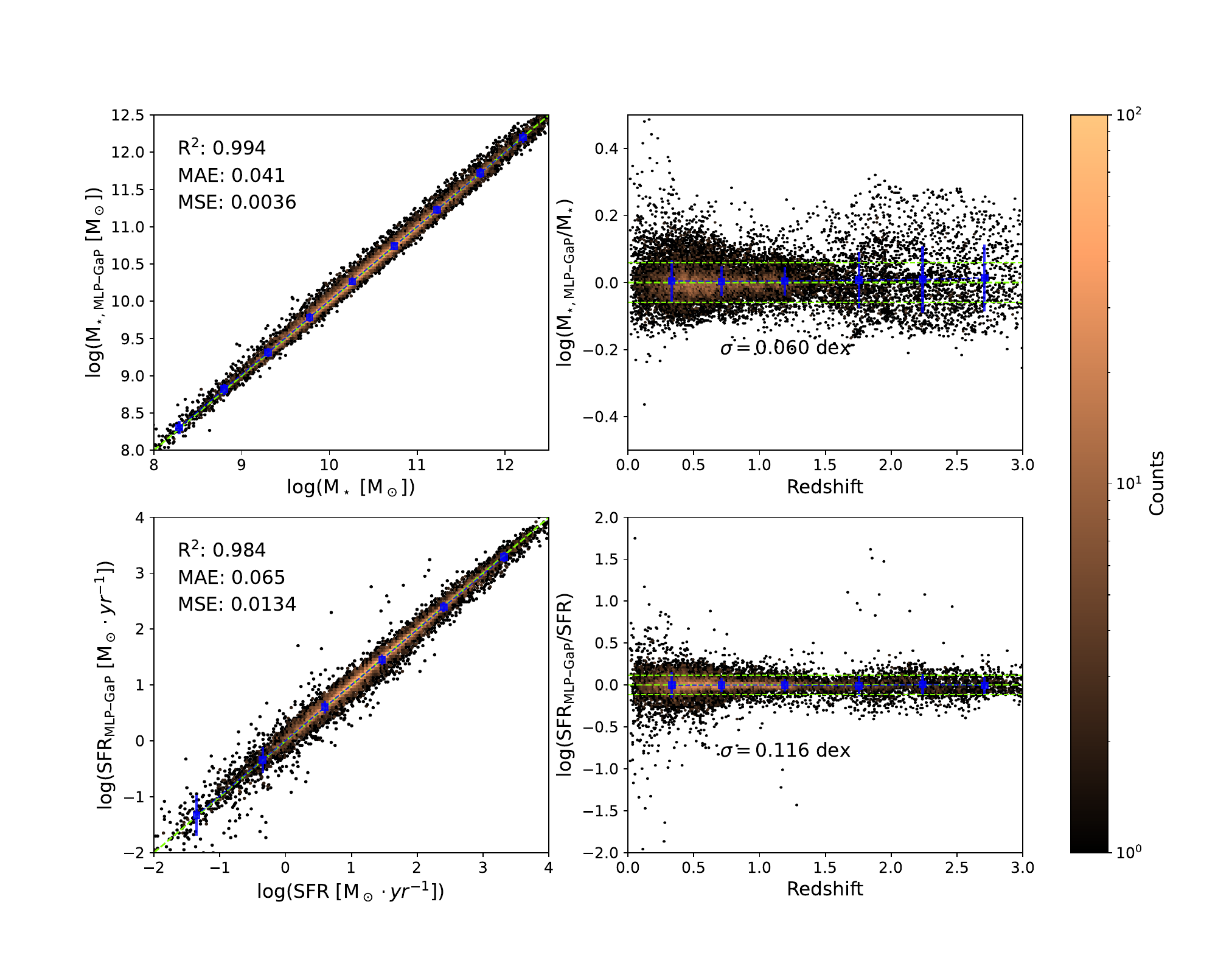}
	\caption{The comparison between the MLP-GaP‘s predictions and reference values in the testing dataset (20,000 galaxies). \emph{Top-Left panel:} The stellar masses of the MLP-GaP are compared with the reference values. \emph{Top-Right panel:} Errors of the MLP-GaP results obtained for the testing dataset as a function of redshift for M$_\star$. \emph{Bottom-Left panel:} The SFRs of the MLP-GaP are compared with the reference values. \emph{Bottom-Right panel:} Errors of the MLP-GaP results obtained for the testing dataset as a function of redshift for SFR.}
	\label{fig:compre-reference}
\end{figure*}
We initially assess the performance of the MLP-GaP using the mock testing dataset. In the top-left panel of Fig.~\ref{fig:compre-reference}, we illustrate the predictions of stellar masses plotted against their reference values. The comparison reveals a strong correlation, with data points closely clustered around the 1:1 line, indicating high accuracy in the predictions. The $R^2$, MAE and MSE are 0.994, 0.041 and 0.0036, respectively, suggesting that the MLP-GaP could accurately predict stellar masses of mock galaxies. To check whether the predictions are affected by the redshifts, in the top-right panel of Fig.~\ref{fig:compre-reference}, we present the variation of the predicted stellar masses, in the form of $\log ({ \mathrm{M}_{\star,\mathrm{MLP-GaP}} / \mathrm{M}_\star})$, against the redshifts. We find that the variation is relatively minor, with an overall standard deviation of $\sigma_{M_\star}=0.060$ dex, which translates to a factor of $10^{\sigma_{M_\star}}=1.15$ relative to the established values.
We also notice that the scatter tends to increase at high redshift, but only with a maximum standard deviation of 0.1 dex, suggesting that the MLP-GaP experiences a decrease in accuracy when predicting stellar masses as redshift increases. Overall, the MLP-GaP performs well in the M$_\star$ prediction in the redshift range of $0<z<3$.

Let's move on to SFR prediction. In our mock sample, some passive galaxies show statistically negligible and very low SFRs, which we can exclude from our analysis. 
We use a standard threshold based on the specific star formation rate (sSFR = SFR/M$_\star$) to filter out these passive galaxies. The threshold we have adopted is $\log{\mathrm{sSFR} >-12}$ \citep[e.g.,][]{2021MNRAS.500.2036K}, which is regarded as the minimum value above which the star formation activity in galaxies cannot be ignored. The following all analyses regarding SFR have filtered out these passive galaxies.
The predicted values of SFRs versus their reference values are plotted in the bottom-left panel of Fig.~\ref{fig:compre-reference}. Similarly to the M$_\star$, most of the points are distributed around the 1:1 line, although with a larger scatter. The evaluation indices for the SFRs indicate a very good accuracy, with a $R^2$=0.984, MAE=0.065, and MSE=0.0134. Although these indices are not as good as those for M$_\star$, they still suggest a high predictive accuracy for SFRs. In the bottom-left corner of this panel, where $\log \mathrm{SFR}<0$, a certain discrepancy in the predicted values is observed, indicating that MLP-GaP's accuracy has declined in the region. In the bottom-right panel of Fig.~\ref{fig:compre-reference}, we plot the variation, $\log{(\mathrm{SFR_{MLP-GaP}/ SFR})}$ of the prediction, against the redshifts. From a general perspective, we can see a relatively minor variation, with a standard deviation of $\sigma_\mathrm{SFR} = 0.116$ dex, which is equivalent to a factor of $10^{\sigma_\mathrm{SFR}} = 1.306$,  relative to established values. 

In summary, our research has established that the MLP-GaP is a robust and accurate tool for predicting the stellar masses and SFRs of galaxies in the testing dataset. It can exhibit consistent performance across a wide range of redshifts.

\subsection{Comparing the MLP-GaP predictions with CIGALE estimations}
To thoroughly evaluate the performance of MLP-GaP, we use CIGALE to estimate the stellar masses and SFRs of the mock testing dataset. This analysis involves fitting multi-band photometric data, and the relevant modules and parameters used in the configuration file of CIGALE are detailed in Table~\ref{tab:module-2}. The results obtained are then compared with those predicted from MLP-GaP to assess the accuracy and precision of the predictions.
\begin{table*}[h]
	\centering
	\caption{The module assumptions for SED fitting.\label{tab:module-2}}
	\begin{tabular}{c|c|c}
		\hline\hline
		Module & Parameters & Value \\
		\hline
		\multirow{5}{*}{SFH(delayed)} & \multirow{2}{*}{Tau\_main(Myr)} & 250, 500, 1000, 1500, 2000, 2500, 3000, 3500, 4000, 4500, 5000,  \\
		&&5500, 6000, 7000, 8000\\
		& \multirow{2}{*}{Age(Myr)} & 250, 500, 1000, 1500, 2000, 2500, 3000, 3500, 4000, 4500, 5000,  \\
		&&5500, 6000, 7000, 8000, 9000, 10000, 11000, 12000, 13000\\
		& f\_burst & 0 \\
		\hline
		\multirow{2}{*}{BC03}&IMF&1(Chabrier)\\
		&Metallicity&0.0001, 0.0004, 0.004, 0.008, 0.02, 0.05\\
		\hline
		\multirow{2}{*}{dustatt\_modified\_starburst} & \multirow{2}{*}{E\_BV\_lines(mag)} & 0.0, 0.01, 0.02, 0.05, 0.1, 0.15, 0.2, 0.25, 0.3, 0.4, 0.5, 0.6, 0.7,  \\
		&&0.8, 0.9, 1.0, 1.1, 1.2, 1.3, 1.4, 1.5\\
		\hline
		dale2014& Alpha& 1.0, 2.0, 3.0, 4.0\\
		\hline
	\end{tabular}
	
\end{table*}

\begin{figure*}
	\centering
	\includegraphics[width=\linewidth]{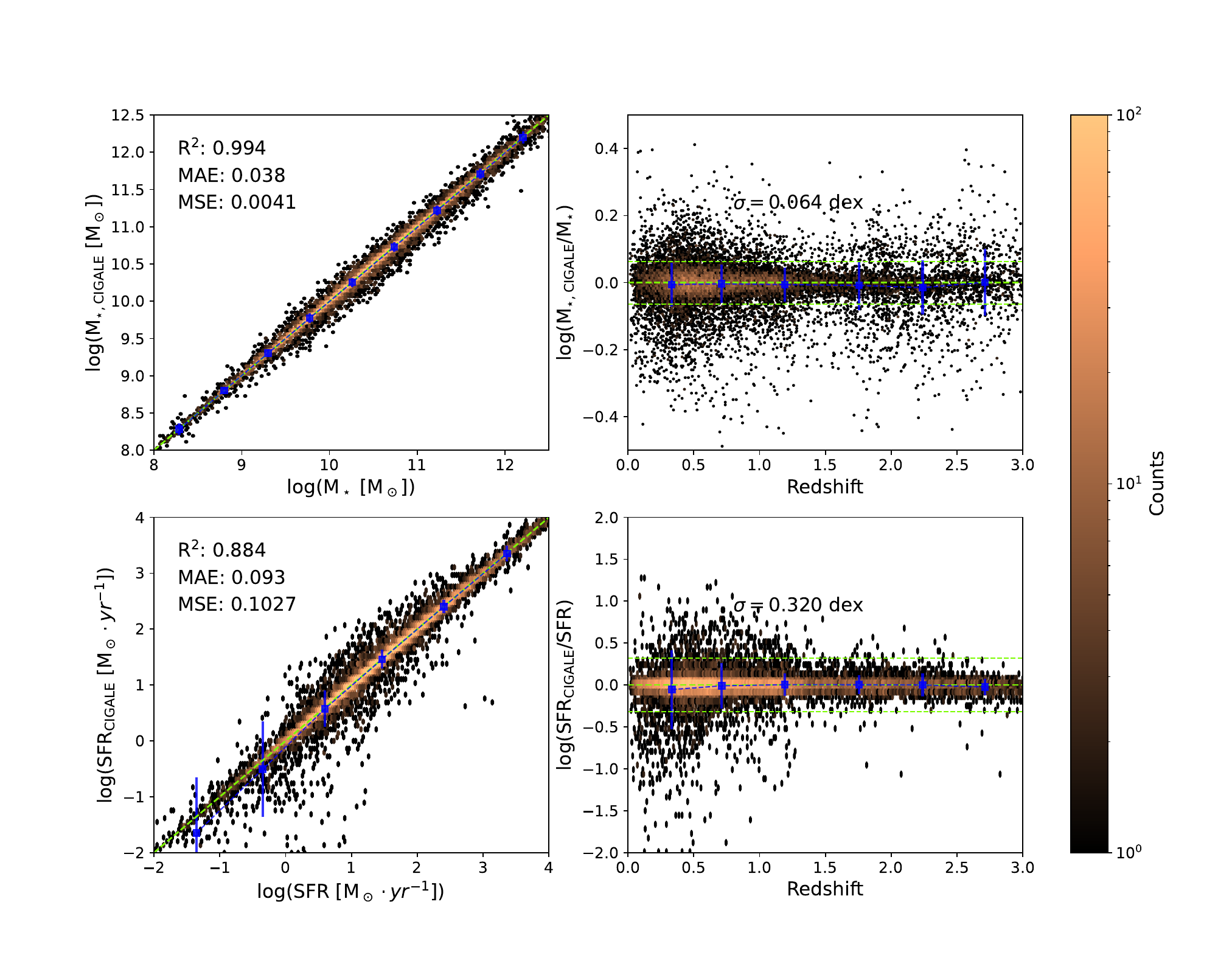}
	\caption{The comparison between the CIGALE predictions and reference values on the testing dataset. \emph{Top-Left panel:} The stellar masses of the CIGALE are compared with the reference values. \emph{Top-Right panel:} Errors of the CIGALE results obtained for the testing dataset as a function of redshift for M$_\star$. \emph{Bottom-Left panel:} The SFRs of the CIGALE are compared with the reference values. \emph{Bottom-Right panel:} Errors of the CIGALE results obtained for the testing dataset as a function of redshift for SFR.}
	\label{fig:compre-CIGALE-reference}
\end{figure*}

\begin{figure*}
	\centering
	\includegraphics[width=\linewidth]{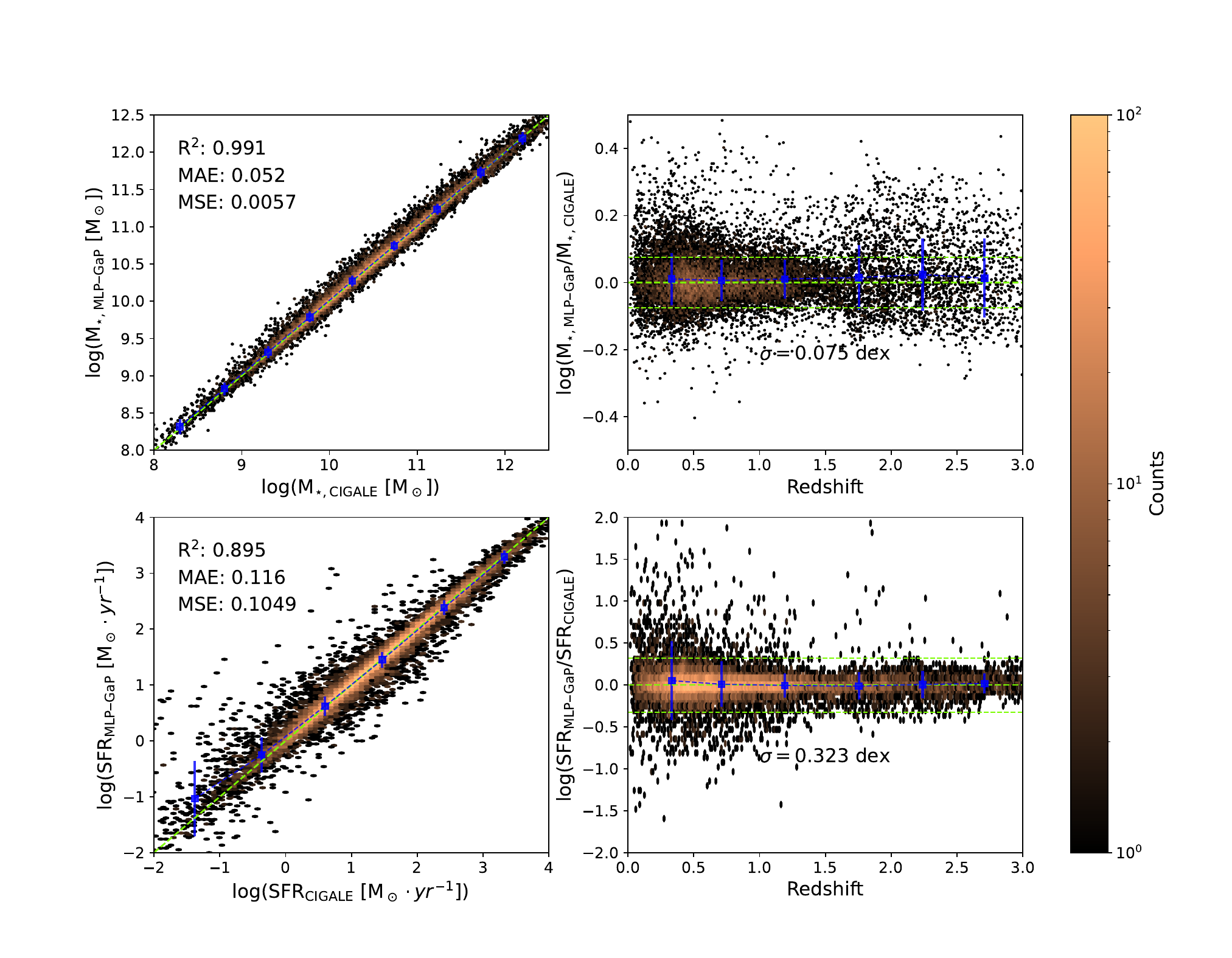}
	\caption{The comparison between the MLP-GaP predictions and CIGALE predictions on the testing dataset. \emph{Top-Left panel:} The stellar masses of the MLP-GaP are compared with those of CIGALE. \emph{Top-Right panel:} Errors between MLP-GaP and CIGALE results obtained for the testing dataset as a function of redshift for M$_\star$. \emph{Bottom-Left panel:} The SFRs of the MLP-GaP are compared with those of CIGALE. \emph{Bottom-Right panel:} Errors between MLP-GaP and CIGALE results obtained for the testing dataset as a function of redshift for SFR.}
	\label{fig:compre-MLP-GaP-CIGALE}
\end{figure*}

In Fig.~\ref{fig:compre-CIGALE-reference}, we present the CIGALE fitting results versus their reference values for both the M$_\star$ (top-left panel) and SFR (bottom-left panel), and their variations as a function of redshifts are also appended in the right panels. In terms of estimating stellar masses, CIGALE is comparable performance to MLP-GaP. This conclusion is supported by the evaluator indices of $R^2$, MAE and MSE, with the deviation being similarly close to that of MLP-GaP. In the top-right panel of Fig.~\ref{fig:compre-CIGALE-reference}, we find that the stellar masses derived using CIGALE exhibit a slight advantage at high redshifts ($z>1.5$). This is evidenced by the relatively smaller error bars within this redshift range, and a greater concentration of data points around the line where $\log{(\mathrm{M_{\star, CIGALE}/M_\star})}=0$. 
Moving on to SFR estimation, the evaluator indices suggest MLP-GaP outperforms CIGALE in estimating SFRs. The standard deviation of the differences between the SFRs derived by CIGALE and the reference values is $\sigma=0.320$ dex, a value that is notably higher compared to the standard deviation of the differences between the SFRs predicted by MLP-GaP and the reference values.
The primary discrepancy arises at low redshifts ($z<1.5$), where CIGALE not only exhibits greater variability in its deviations but also deviates from the line representing $\log{(\mathrm{SFR_{CIGALE}/SFR})}=0$. 
These findings suggest that MLP-GaP can provide more accurate predictions of SFRs for mock galaxies in the testing dataset compared to CIGALE.

Subsequently, a direct comparison will be made between the predicted values from MLP-GaP and those derived using CIGALE. Such a comparison is pivotal because, in actual astronomical observations, the ``ground truth'' values of the stellar population parameters are unattainable. All values are dependent on various assumptions, models, and inherent uncertainties, which means that error is an inevitable aspect of the process. Therefore, the performance of a novel tool like MLP-GaP must be evaluated by comparing its results with those obtained from standard tools or methodologies.
In the left panels of Fig.~\ref{fig:compre-MLP-GaP-CIGALE}, the vertical axis represents the results from MLP-GaP, while the horizontal axis displays the corresponding results from CIGALE. The majority of sources are concentrated near the diagonal line, indicating a good consistency between the predictions made by the two tools. Additionally, the results from their evaluators also demonstrate this point.
The right panels of Fig.~\ref{fig:compre-MLP-GaP-CIGALE} illustrate the deviation of their results as a function of redshifts. The top-right panel indicates that the M$_\star$ estimates derived from both MLP-GaP and CIGALE are in close alignment, exhibiting a standard deviation of $\sigma=0.075$ dex. 
The bottom-right panel also suggests that the SFRs estimated from MLP-GaP and CIGALE agree, with a standard deviation of $\sigma=0.323$ dex. 
These mean that MLP-GaP is capable of delivering predictions that align with those of standard tools, demonstrating its reliability and potential as a viable alternative for estimating stellar masses and star formation rates in galaxies.

Finally, an evaluation is conducted on the computational efficiency of MLP-GaP compared to the traditional standard tools. The two primary indicators for assessing computational efficiency are the time consumption and the utilization of computational resources. To ensure a fair and unbiased comparison, both CIGALE and MLP-GaP are run on identical hardware platforms. Specifically, the tests are run on a computing system equipped with an Intel Core i7-11700F processor, featuring 12 cores operating at a base frequency of 2.5GHz. For MLP-GaP, the estimation of stellar masses and SFRs for 20,000 galaxies is conducted using a single core, with the entire process being completed in 11.018 seconds. In contrast, despite utilizing 10 cores, the SED fitting method still requires a longer computational time, approximately 200 minutes. When not considering the number of cores used, the time expended by the SED fitting method is already 985 times greater than that of MLP-GaP. Should MLP-GaP also use 10 cores, it would undoubtedly achieve an even more rapid execution speed. Therefore, MLP-GaP demonstrates outstanding performance in terms of running speed, showcasing its potential as a highly efficient tool for astronomical data analysis.

In conclusion, MLP-GaP demonstrates a superior ability to predict the stellar masses and SFRs of galaxies with higher precision in comparison to CIGALE. Furthermore, its computational performance is exceptionally impressive, providing a remarkable running speed that substantially surpasses that of traditional tools. 


\section{Discussion}\label{sec:discussion}
\subsection{Performance on actual dataset}\label{sec:real_data}
Although the mock dataset closely approximates actual observational data, inherent differences may still exist, particularly in the distribution of parameters and the interrelationships among them. Therefore, relying solely on the mock dataset to test MLP-GaP may not fully confirm its reliability. To demonstrate the science-readiness of the MLP-GaP, we will advance to assess its performance using actual observational data. The catalog provided by \cite{2023SCPMA..6629513X} is highly suitable for serving as a test dataset for MLP-GaP. This catalog includes the observational data for 288,809 galaxies and the stellar population parameters for their galaxies as outputted by LePhare and CIGALE. Here, we will apply MLP-GaP and CIGALE to the catalog to estimate the stellar masses and SFRs of the galaxies, and then compare these estimations.
The redshifts used in this comparison are ``morphoto-Z'' obtained by GaZNet \citep{2022A&A...666A..85L}.

\begin{figure*}
	\centering
	\includegraphics[width=\linewidth]{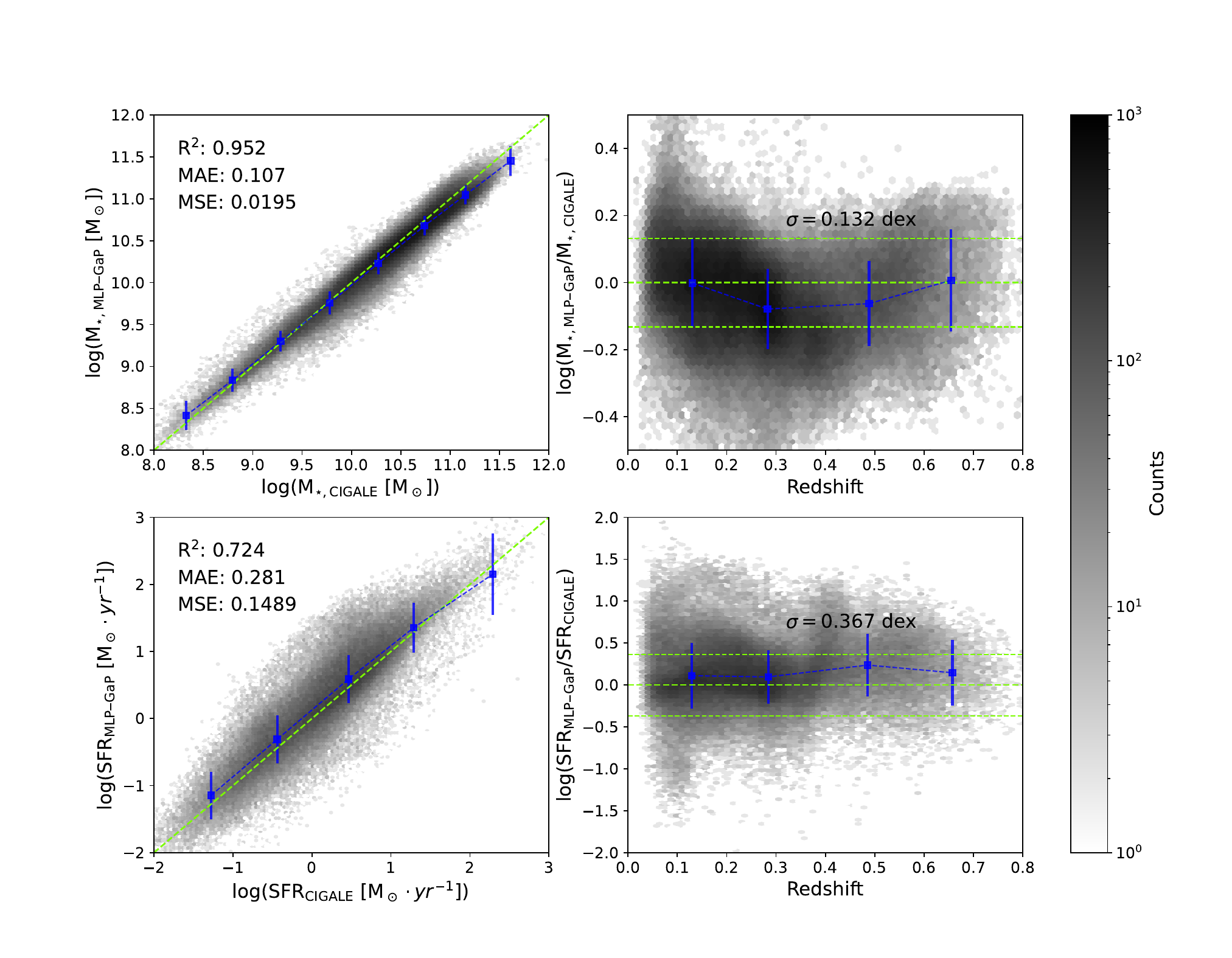}
	\caption{The comparison between the MLP-GaP predictions and CIGALE predictions on the actual dataset. \emph{Top-Left panel:} The stellar masses of the MLP-GaP are compared with those of CIGALE. \emph{Top-Right panel:} Errors between MLP-GaP and CIGALE results obtained for the testing dataset as a function of redshift for M$_\star$. \emph{Bottom-Left panel:} The SFRs of the MLP-GaP are compared with those of CIGALE. \emph{Bottom-Right panel:} Errors between MLP-GaP and CIGALE results obtained for the testing dataset as a function of redshift for SFR.}
	\label{fig:compre-X23}
\end{figure*}

Fig.~\ref{fig:compre-X23} presents a detailed comparison of the performance of MLP-GaP and CIGALE on actual observational data of galaxies. The top-left panel demonstrates a good consistency in the stellar masses estimation between MLP-GaP and CIGALE, as evidenced by $R^2=0.952$, MAE=0.107, MSE= 0.0195. The top-right panel indicates that the deviations in their stellar masses exhibit some variation with redshift, but these fluctuations are minor than their standard deviation $\sigma =0.132$ dex. Compared to the results from the testing dataset, the consistency between MLP-GaP and CIGALE in estimating the stellar masses of actual galaxies has slightly deteriorated. 
The bottom-left panel illustrates the agreement between the SFR estimates by MLP-GaP and CIGALE, albeit with a certain degree of dispersion. The bottom-right panel demonstrates that variations in SFR estimation exhibit some redshift dependency, yet these fluctuations are relatively minor compared to the standard deviation. In estimating the SFRs of actual galaxies, there is also some degradation in consistency.
This may arise from inherent limitations within MLP-GaP itself. Given that the training data is mocked, there are certain discrepancies in the distribution of parameters compared to actual galaxies. For instance, the fraction of massive galaxies in both the training and testing datasets may be significantly higher than that found in actual galaxies. This could lead MLP-GaP to learn patterns that do not align with those of actual galaxies, particularly in the case of massive mock galaxies. 
Regardless, MLP-GaP still demonstrates good consistency with CIGALE in estimating stellar masses and SFRs of actual galaxies. Therefore, MLP-GaP can serve as an alternative to traditional SED fitting tools for predicting stellar masses and SFRs. In particular, MLP-GaP is significantly faster than conventional methods in terms of computation speed, so it is more suitable for estimating the stellar masses and SFRs of billions of galaxies in large-scale surveys.

\subsection{Application and improvement in future}\label{sec:further_improvement}

Shortly, astronomy will enter a new era of development, where an unprecedented wealth of observational data can be obtained from various large-scale and deep-area surveys. This new age will be marked by the availability of data from major projects. 
These surveys cover vast sky areas, providing multi-band (UV, optical, and NIR) photometric data and images for billions of galaxies, thus presenting unique opportunities for scientific inquiry. Our MLP-GaP, leveraging the power of ML, offers significant advantages over traditional SED fitting techniques. It is uniquely positioned to swiftly and accurately estimate the stellar masses and SFRs of the galaxies observed in these large-scale surveys. As the volume of observational data continues to expand, the implementation of efficient and accurate ML algorithms is poised to become increasingly invaluable. The capability of MLP-GaP to rapidly and accurately predict the physical parameters for billions of galaxies expands our comprehension of the cosmos, unveiling new perspectives on the evolution of galaxies throughout the universe, thereby significantly advancing our knowledge of the astrophysical processes that shape the cosmos.

While MLP-GaP has demonstrated its potential in predicting stellar masses and SFRs for huge volumes of galaxies, it requires further and profound enhancements to fully realize its capabilities. Our roadmap for future improvements includes the following aspects: 
\begin{enumerate}
	\item {\bf Enhancing Training Data Diversity:}
	Our current MLP-GaP is trained on mock datasets generated by CIGALE. Moving forward, we want to achieve that the parameter distributions of mock galaxies closely mirror those of actual galaxies.  Moreover, we intend to enrich the training datasets with more intricate galaxy formation histories, a wider array of dust attenuation models, and different stellar population models. Additionally, we will integrate various astrophysical processes and observational noise to better simulate real-like data, thereby enhancing the generalizability and accuracy of the MLP-GaP.
	\item {\bf Optimizing Model Architecture:}
	Although the MLP model utilized in this study has shown its power, there is ample scope for optimization. Our future endeavours will not only seek to augment the model's computational speed but also refine its predictive accuracy. We will investigate alternative network architectures, including Self-Attention and Transformer Models, which have demonstrated remarkable performance in various domains.
	\item {\bf Expanding Parameter Prediction:}
	We aim to extend MLP-GaP to predict more parameters of galaxies. This will encompass characteristics such as age, metallicity, IMF, and so on, providing a more comprehensive understanding of the properties of galaxies.
	\item {\bf Uncertainty Quantification:}
	In the realm of scientific inquiry, the accurate estimation of parameter uncertainties is critical for robust error analysis. Recognizing this, we will adopt Bayesian methods to determine the uncertainties of different parameters. By integrating Bayesian networks within our ML algorithm, we can better quantify the uncertainties associated with our predictions. It will provide not only point estimates of the parameters but also their probabilistic distributions. Such an approach is essential for credible scientific discourse and for making informed decisions in the light of inherent uncertainties in observational data.
	
\end{enumerate}

\section{Summary} \label{sec:summary}
In the era of large scale surveys, there is a massive amount of observational data available. However, the challenge remains of how to rapidly and accurately derive various stellar population parameters for billions of galaxies.
Among the numerous parameters that characterize galaxies, the M$_\star$ and SFR are considered the most critical. In response to the aforementioned challenge, we have developed an ML algorithm called MLP-GaP, which can rapidly and accurately predict the stellar masses and SFRs for a massive amount of galaxies. 

Firstly, we used CIGALE to generate a mock dataset, which is a catalog consisting of 120,000 mock galaxies. This catalog provided redshift, M$_\star$, SFR, and photometric data of nine bands for each mock galaxy. The mock dataset was meticulously partitioned into three distinct subsets, with a training dataset comprising 90,000 galaxies, a validation dataset of 10,000 galaxies, and a testing dataset of 20,000 galaxies. Subsequently, we used an MLP model with 10 layers to build MLP-GaP. Through rigorous training and validation processes, MLP-GaP was optimized to yield predictions that were consistent with the reference values on the testing dataset.
Furthermore, MLP-GaP demonstrated a significant faster in processing speed compared to CIGALE. To demonstrate the science-readiness of the MLP-GaP, we applied to actual galaxy samples. The predicted values from MLP-GaP exhibited a commendable level of consistency with the estimated values derived using SED fitting. 
This consistency suggested MLP-GaP could serve as an alternative to traditional SED fitting tools for predicting stellar masses and SFRs. Given the outstanding processing speed of MLP-GaP, it can be considered an essential tool for estimating the parameters of billions of galaxies in the era of large scale surveys.


\normalem
\begin{acknowledgements}
We sincerely thank the anonymous reviewer for useful suggestions.
We acknowledge the support of \emph{National Nature Science Foundation of China} (Nos  12303017, 12203096).
This work is also supported by \emph{Anhui Provincial Natural Science Foundation} project number 2308085QA33.
This work is supported by the science research grants from the China Manned Space Project.

\end{acknowledgements}
  
\bibliographystyle{raa}
\bibliography{bibtex}

\end{document}